\documentclass{kluwer}    
\usepackage{epsfig}

\newdisplay{guess}{Conjecture}

\begin{document}                                                                                   
\begin{article}
\begin{opening}         
\title{Common Proper Motion Search for Faint Companions 
Around Early-Type Field Stars - Progress Report} 
\author{Valentin D. \surname{Ivanov}, G. Chauvin, C. Foellmi, 
   M. Hartung, N. Hu\'elamo, C. Melo, D. N\"urnberger and 
   M. Sterzik}  

\runningauthor{V.D. Ivanov}
\runningtitle{Faint Companions Around Early-Type Field Stars}
\institute{European Southern Observatory, Ave. Alonso de
   Cordova 3107, Casilla 19, Santiago 19001, Chile}
\date{September 30, 2005}

\begin{abstract}
The multiplicity of early-type stars is still not well
established. The derived binary fraction is different for 
individual star forming regions, suggesting a connection with 
the age and the environment conditions. The few studies that 
have investigated this connection do not provide conclusive 
results. To fill in this gap, we started the first detailed 
adaptive-optic-assisted imaging survey of early-type field 
stars to derive their multiplicity in a homogeneous way. The 
sample has been extracted from the Hipparcos Catalog and 
consists of 341 BA-type stars within $\sim$300\,pc from the 
Sun. We report the current status of the survey and describe 
a Monte-Carlo simulation that estimates the completeness of our 
companion detection.
\end{abstract}
\keywords{stars:binaries:general, stars:binaries:visual, 
stars:early-type}

\end{opening}           

\section{Introduction}

The multiplicity of pre-main sequence (PMS) and main sequence
(MS) late-type stars have been extensively studied: Duquennoy
\& Mayor \cite{duq91}, Reipurth \& Zinnecker \cite{rei93}, Prosser
et al. \cite{pro94}, Brandner et al. \cite{bra96}. PMS late-type 
stars in low-density clouds like Taurus show higher binary 
fractions than PMS late-type stars formed in massive and dense 
star forming regions (SFR) like Orion. This difference was 
explained with an environmental dependence of the binary fraction: 
low-mass stars born in dense SFRs show a higher probability of 
dynamic interactions with more massive members, so that they are 
ejected resulting in a low binary fraction, as shown by Sterzik \& 
Durisen \cite{ste95}. On the other hand, stars in low-density SFRs 
undergo low rate of stellar encounters, resulting in a higher 
binary fraction.

It is unclear if this result remains valid for early-type stars,
mostly because of the difficult detection of faint companions near
bright O- and B-type stars. The works of Petr et al. \cite{pet98}, 
Preibisch et al. \cite{pre99} and Shatsky \& Tokovinin \cite{sha02} 
indicate that the binary fraction of early-type stars varies from 
one SFR to another but current statistical basis is still not solid 
enough. The age of the SFRs can also affect the binary statistics: 
older systems are expected to have undergone through more dynamical 
interaction, reducing their binarity fraction in comparison with 
the younger ones, as point by Tokovinin et al. \cite{tok99}.
                                                                                
\section{The Survey}

The cluster multiplicity studies can cover only a limited range
of density and age. This prompted us to estimate the binarity
fraction of a representative, volume-limited sample of early-type
field stars. We designed a survey able to detect at
$\sim$10$\sigma$ level an M4-type companion at the mean distance
of our sample ($\sim$200 pc) down to
0.4 arcsec separation from 100\,Myr old A-type primary. The
companions around B-type stars will be younger ($\sim$10 Myr),
brighter, and easier to detect. Last but not least, the physical
nature of the candidate components is verified by their common
proper motion. Our goal is to compare the properties such as
incidence and mass ratio of the multiple stars in the field and
in different star-forming regions.

\section{Sample Selection}

The sample stars were selected according to the following criteria:

\vspace{1mm}
$\bullet$ Apparent color $B$$-$$V$$\leq$0.2\,mag - conservative 
criterion met by all unreddened BA stars, and a few contaminating 
later-type stars. The spectral types for all stars were verified 
to be BA.

$\bullet$ The sample contains field stars only. Members of the
OB-associations listed in de Zeeuw et al. \cite{dez99} were 
excluded.

$\bullet$ Distance D$\leq$300\,pc from the Sun (HIPPARCOS). At 
D=300\,pc the telescope's diffraction limit of 0.07\,arcsec 
corresponds to 21\,A.U (Shatsky \& Tokovinin \cite{sha02} probed 
separations 45-900\,A.U).

$\bullet$ Proper motions $\geq$27\,mas/yr allowing to confirm/reject 
physical companion candidates taking two epoch separated by 
1-2\,yr.

$\bullet$ The apparent $V$=5-6\,mag, so the targets are suitable 
self-references for NACO even under poor weather conditions.

$\bullet$ The stars have DEC$\leq$0\,deg, i.e. visible from the VLT.
\vspace{1mm}

The final sample contains 341 field B- and A-type stars. The 
average distance is 114\,pc, the median distance is 104\,pc.

\section{Observations and Current Status}

The observations were carried out with NAOS--CONICA (Nasmyth 
Adaptive Optics System -- Near-Infrared Imager and Spectrograph) 
at the ESO VLT over the last two years. The pixel scale was 
27.03\,mas\,px$^{-1}$, giving 27.7$\times$27.7 arcsec field 
of view. Each target was observed at 9 different position on the 
detector, collecting total of $\sim$7.5\,min of integration. The 
data reduction includes sky subtraction, flat-fielding, 
aligning and combination of the images into a single frame. 
Next, we perform PSF fitting/subtraction and search for faint 
companions. 

As of Sept 2005 we have observed 196 objects from our sample. We 
have analyzed 152 of them: 81 appear single, and 71 show companion 
candidates whose nature will be tested with the second epoch 
observations with $\geq$2\,yr baseline.

\section{Analysis: Modeling the Survey}

\begin{figure}
\centerline{\epsfig{file=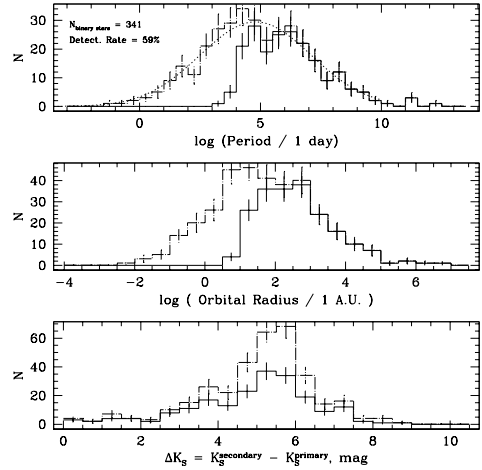,width=16pc}}
\caption{Monte-Carlo simulation of the survey completeness:
orbital period distribution (top panel), orbital radii 
distribution (middle), and primary-secondary $K$-band 
magnitude difference (bottom). 
The dot-dashed lines show the ``true'' adopted/generated 
distributions and the solid lines are the ``observed'' 
distributions after the observational detectability 
conditions have been applied.
The theoretical period distribution of Duquennoy \& Mayor 
(1991) is also shown (doted line on the top panel).}
\label{Model}
\end{figure}

To estimate the sensitivity and the completeness of the survey we
have created a Monte-Carlo simulation that takes into account all
available information for the survey stars (Fig.\,\ref{Model}). 
The model input parameters are:

\vspace{1mm}
$\bullet$ the known distances, spectral types, absolute 
luminosities for all primaries;

$\bullet$ adopted binarity fraction;

$\bullet$ secondary star mass – randomly sampled from the Kroupa 
et al. \cite{kro93} IMF;

$\bullet$ secondary star's spectral type and absolute magnitude – 
calculated from the mass;

$\bullet$ orbital periods - randomly generated from Duquennoy \& 
Mayor \cite{duq91} distribution;

$\bullet$ major axis – calculated from the Kepler's low and the 
period;

$\bullet$ random ellipticity, random orbital inclination;

$\bullet$ visibility criterion based on the magnitude difference 
and the angular separation between the primary and the companion.
\vspace{1mm}

The model predicts: the distributions of periods, angular 
separations, magnitude differences and spectral types for the 
detected binaries. The simulations indicate that we will detect 
about 2/3 of the physical companions.

\acknowledgements
We are grateful to our colleagues from the ESO-Paranal Science 
Operations Department who carried out these observation in 
Service Mode.

\end{article}
\end{document}